\documentclass[10pt,twocolumn,twoside]{IEEEtran}

\usepackage{graphicx,epsfig}
\usepackage{cite}
\usepackage{stfloats}
\usepackage{amsmath}
\usepackage{amsfonts,helvet}
\usepackage{caption}
\usepackage{subcaption}
\usepackage{flushend}
\usepackage{subfig}
\usepackage{hhline}

\newtheorem{theorem}{Theorem}

\newtheorem{algo}[theorem]{Algorithm}

\newtheorem{proposition}[theorem]{Proposition}

\newcounter{proposition}
\begin{document}
\title{Channel and Noise Models for Nonlinear Molecular Communication Systems}


\author{\authorblockN{ Nariman Farsad,$^{\dag \ddag}$~\IEEEmembership{Student~Member,~IEEE}, Na-Rae Kim,$^{\ast \ddag}$~\IEEEmembership{Student~Member,~IEEE}, \\  Andrew W. Eckford,$^\dag$~\IEEEmembership{Member,~IEEE}},  Chan-Byoung Chae,$^\ast$~\IEEEmembership{Senior~Member,~IEEE}

\thanks{$^{\ddag}$ The authors have equally contributed. $^{\ast}$N.-R. Kim and C.-B. Chae are with the School of Integrated Tech., Yonsei University, Korea. Email: \{nrkim, cbchae\}@yonsei.ac.kr. $^{\dag}$N. Farsad and A. Eckford are with Dep. of Computer Science and Engineering, York University, Canada. Email: nariman@cse.yorku.ca, aeckford@yorku.ca. This work was in part supported by the MSIP (Ministry of Science, ICT $\&$ Future Planning), Korea in the ICT R$\&$D Program 2013 (NIPA-2013-H0203-13-1002), a Discovery grant from the Natural Sciences and Engineering Research Council (NSERC) in Canada.}}



\maketitle \setcounter{page}{1} 

\begin{abstract}
Recently, a tabletop molecular communication platform has been developed for transmitting short text messages across a room. The end-to-end system impulse response for this platform does not follow previously published theoretical works because of imperfect receiver, transmitter, and turbulent flows. Moreover, it is observed that this platform resembles a nonlinear system, which makes the rich body of theoretical work that has been developed by communication engineers not applicable to this platform. In this work, we first introduce corrections to the previous theoretical models of the end-to-end system impulse response based on the observed data from experimentation. Using the corrected impulse response models, we then formulate the nonlinearity of the system as noise and show that through simplifying assumptions it can be represented as Gaussian noise. Through formulating the system's nonlinearity as the output a linear system corrupted by noise, the rich toolbox of mathematical models of communication systems, most of which are based on linearity assumption, can be applied to this platform.  
\end{abstract}




\section{Introduction}
\label{Sec: Main}

Today modern telecommunication systems transfer information through use of electrical or electromagnetic signals. There are, however, still many applications where these technologies are not convenient or appropriate. For example, use of electromagnetic wireless communication inside networks of tunnels, pipelines, or unpredictable underwater environments, can be very inefficient. As another example, at extremely small dimensions, such as between micro- or nano-scaled robots~\cite{sen12}, electromagnetic communication is challenging because of constraints such as the ratio of the antenna size to the wavelength of the electromagnetic signal~\cite{aky08}. 

Inspired by nature, one possible solution to these problems is to use {\em chemical signals} as carriers of information, which is called {\em molecular communication}~\cite{hiy05,eckBook}. In molecular communication, a transmitter releases small particles such as molecules or lipid vesicles into a fluidic or gaseous medium, where the particles propagate until they arrive at a receiver. The receiver then detects and decodes the information encoded in these particles. Messages can be encoded in the different properties such as concentration~\cite{mah10,Kuran_ICC11}, number~\cite{far11NanoCom,far12NanoBio}, type~\cite{cob10}, release timing~\cite{eck09} , and/or ratio~\cite{JSAC_Kim12} of molecules. The information-carrying molecules that are released by the transmitter can propagate through different means such as, active transport using molecular motors~\cite{eno11,hiy10LabChip, far12NANO}, diffusion~\cite{nak08, JSAC_10, kur12, pie13}, flow~\cite{TIT_Andrew12, far12NanoBio}, and bacteria~\cite{gre10, cob10, lio12} until they arrive at the receiver, where they are detected and the intended message is decoded.

Molecular communication has several advantages over traditional wireless communication, with the most notable ones being scalability and energy efficiency. For example, in nature, molecular communication is used for intra/inter-cellular communications at the micro or nanoscales~\cite{BIOCHEM}, and used as pheromones for communication between the same species at the macroscale~\cite{ago92}. Moreover, these systems consume much less energy compared to radio based communication systems~\cite{BIOCHEM, kur10}. Finally, molecular communication can be biocompatible and can be manipulated at nano or molecular scales, which makes it ideal for biomedical applications~\cite{ata12}. There are also many potential applications for molecular communication at macroscopic scales, such as communication inside city infrastructure, and robotic communication~\cite{far13}.

Despite all the recent work on molecular communication no practical platform for demonstrating the feasibility of molecular communication has been developed at either the microscale or the macroscale. Recently in~\cite{far13}, the first tabletop platform capable of transporting short text messages across a room using molecular communication was developed. This platform was purposefully designed to be inexpensive and simple such that other researchers could use it as an experimental tool. As demonstrated in~\cite{far13}, the end-to-end system impulse response for this platform tends to be nonlinear. Most communication-theoretic models are based on the assumption that the underlying systems are linear~\cite{pie10}. Therefore, many techniques applied to these systems (e.g. multiple-input multiple-output), can not be implemented on the platform before investigating and modelling this nonlinearity.

\begin{figure*}[t]
	\begin{center}
       \includegraphics[width=\textwidth]{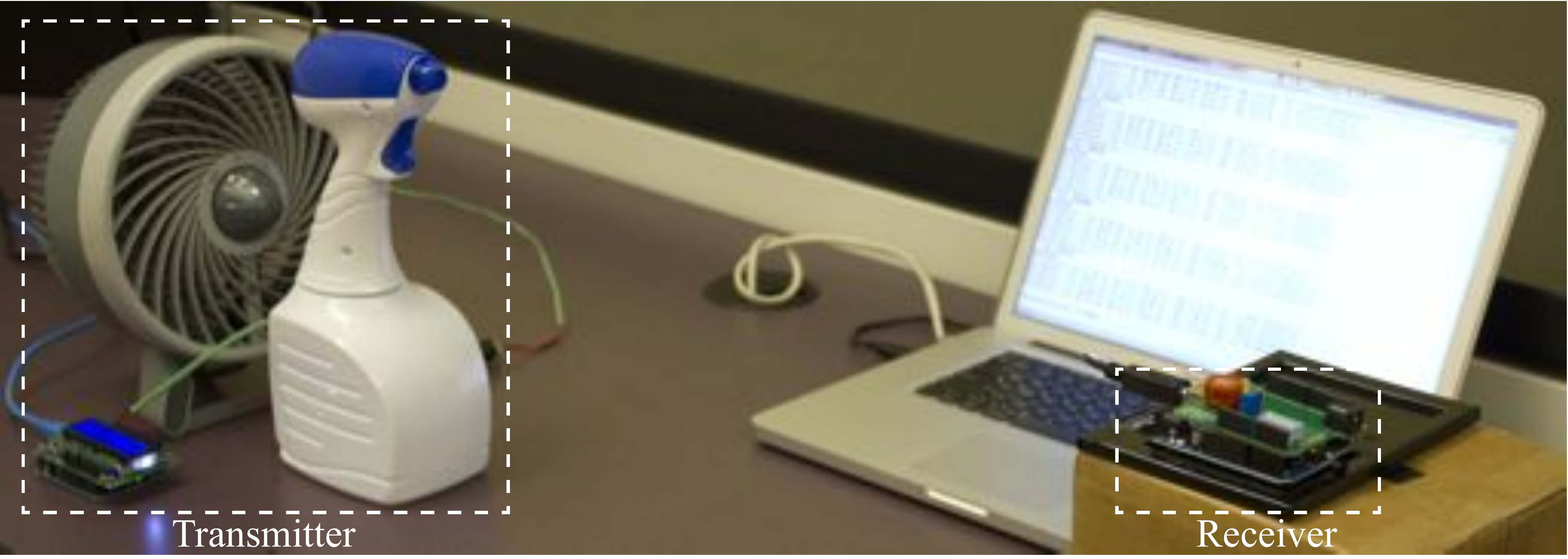}
	\end{center}
    \caption{A tabletop test bed setup for molecular communication.}
    \label{Fig:siso}
\end{figure*}
Another issue that is observed in this platform is the difference between the theoretical system response based on previous works and the observed system response from experiments. Although the exact reason for this discrepancy is not known, some likely causes include the imperfect receiver (i.e. sensor), turbulent flows, and other environmental factors such as random flows within the room. In this work, we first find new mathematical models for the end-to-end system impulse response based on experimental results, and show that the test bed's system response can be modelled fairly accurately with some corrections to the previously published theoretical models. 

After deriving these corrected models, we investigate the nonlinearity of the platform. In particular, we show that the nonlinearity can be modelled as noise. Although the exact cause of nonlinearity is not known, some possible culprits include receiver imperfections, transmitter imperfections, or turbulent flow. For example, the spray that is used for releasing the chemicals does not produce consistently-sized droplet in the spray stream across different trials. Moreover, the sensor is prone to response and recovery times~\cite{boc10}. Despite this inherent nonlinearity, we show that a nonstationary noise process, which can be reduced to a Gaussian noise through simplifying assumptions~\cite{sto11, par12}, can be used to model this nonlinearity. Therefore, the nonlinearity present in the platform can be effectively modelled as the output of a linear system corrupted by noise. We apply the derived model to the experimental measurements obtained from the tabletop platform and the results indicates that the noise model is an effective tool for representing the nonlinearity. Therefore, the rich body of theoretical work that has been developed in the past by communication engineers can be applied to this platform.

The rest of this paper is organized as follows. Section II investigates the system model under consideration explaining details about the test bed and experimental setup. Section III and IV proposes a new mathematical model and nonlinearity analysis of the system, respectively. We conclude the paper in Section V.

\begin{figure}[t]
 	\begin{center}
		\includegraphics[width=\columnwidth]{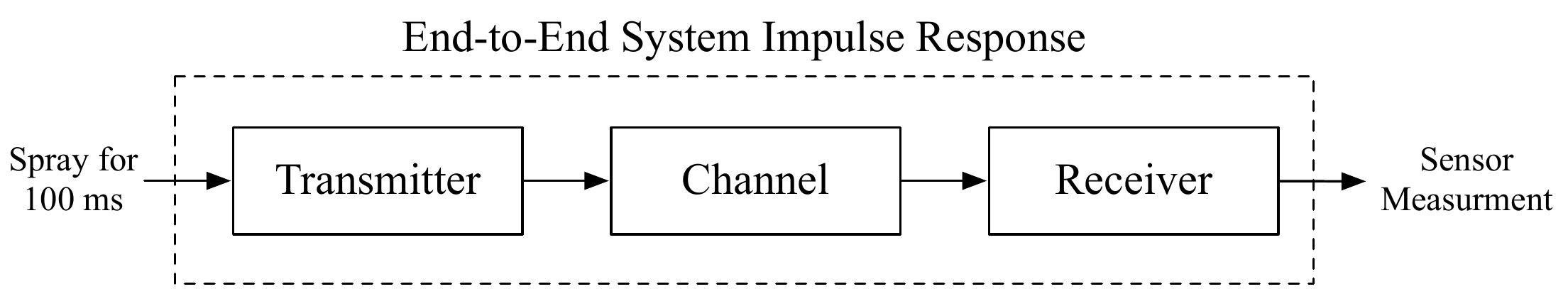}
	\end{center}%
    \caption{The end-to-end system impulse response is generated by using a very short spray.}
    \label{Fig:impResBlock}
\end{figure}

\section{Experimental Setup and Previous Theoretical Models}
\label{Sec:Model} 

\subsection{Tabletop Test bed}
\label{Subsec:Testbed}

The macroscale tabletop test bed which was presented in~\cite{far13}, and is used for the experiments in this paper, is shown in Fig. \ref{Fig:siso}. The transmitter is composed of a spray for releasing the signalling molecules, a fan for assisting the propagation, and a microcontroller with an LCD display and push buttons for controlling the spray. When an input is given to the microcontroller, the information is converted into a binary stream, which can be transmitted through different modulation schemes by precisely controlling the spray.

The signalling chemical that is used for transmission of information in this paper is isopropyl alcohol (one isomer of propanol with the molecular formula of C$_3$H$_7$OH).
When the spray releases these molecules, they propagate through the medium (i.e. air), assisted by the flow generated by the fan behind the spray. In this work, we use the Honeywell 7 inch Personal Tech Fan to create flows in the medium. 

The receiver consists of an alcohol sensor and a microcontroller that reads the sensor data. Since isopropyl alcohol is used as carrier of information, MQ-3~\cite{MQ3} semiconducting metal-oxide gas sensor is used for detection at the receiver. This sensor can measure the concentration of different types of alcohol. The microcontroller at the receiver side reads the sensor data using an analog to digital converter. The data can then be analyzed and sent to a computer through serial port. In~\cite{far13}, it was shown that short text messages could be transmitted across a room using this setup through simple on-off keying (also known as concentration shift keying~\cite{kur11}). It was also shown that the system tends to be nonlinear. In this work, we first analyze the system response of the platform more closely, and derive theoretical models for this test bed.  We then use the derived theoretical models to investigate the nonlinearity property of the system in great details.

The end-to-end system impulse response for this platform can be obtained by using a very short spray (e.g. 100 ms) at the transmitter, which resembles the delta function from signal processing, and measuring the sensor output at the receiver. This is demonstrated in Fig.~\ref{Fig:impResBlock}. The end-to-end system impulse response includes the transmitter block, the channel and the propagation mechanise, and the receiver block. Therefore, the effects of all these blocks are incorporated in the end-to-end system impulse response. 

\begin{figure}[!t]
 \centerline{\resizebox{\columnwidth}{!}{\includegraphics{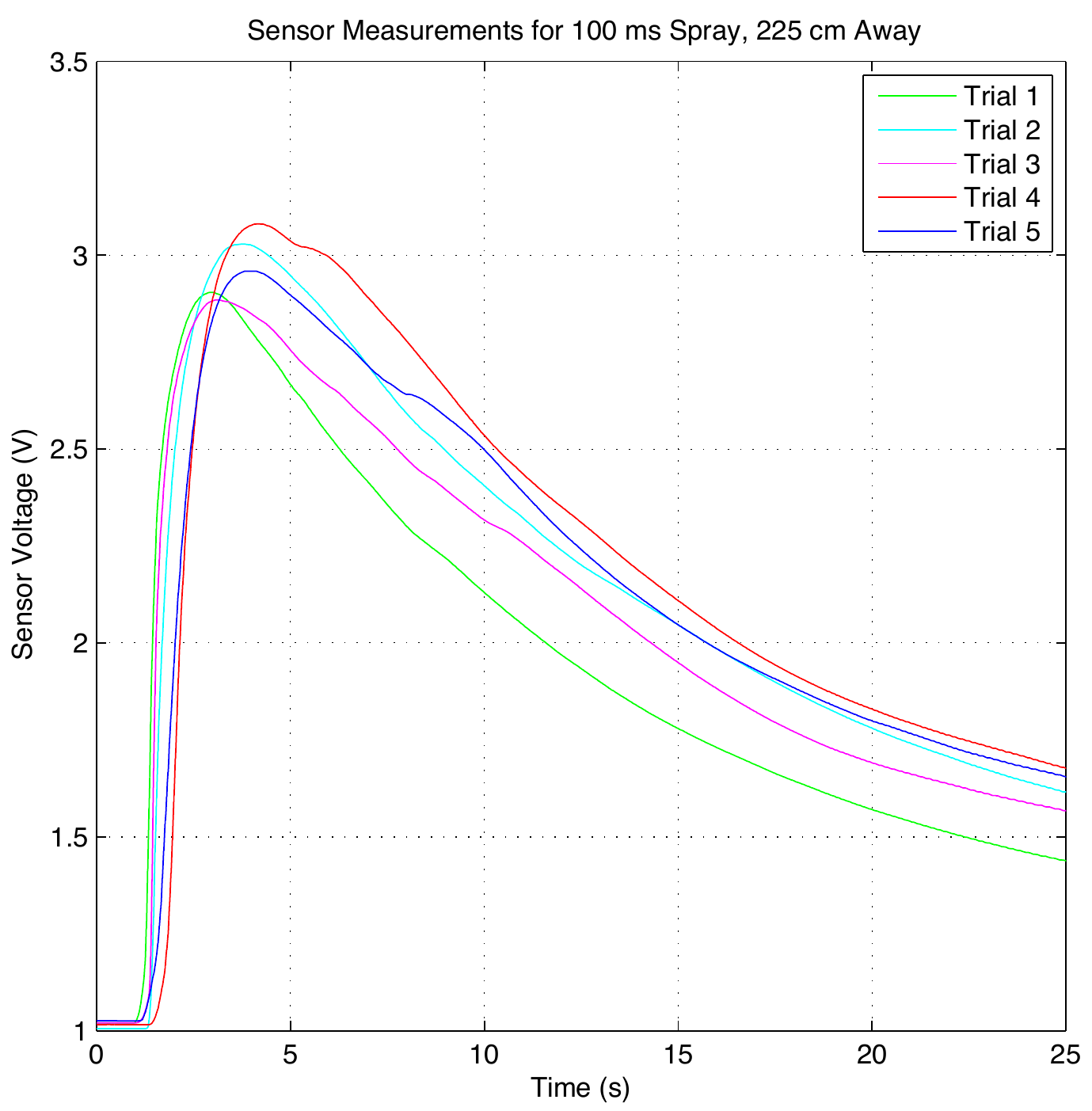}}}
  \caption{The end-to-end system impulse response obtained experimentally across five different trials.}
  \label{Fig:sysResDifTrial}
\end{figure}

To perform some measurements, we separate the transmitter and the receiver by 225 cm. We select 225 cm as an example, and the separation can be any distance as shown in~\cite{far13}. At the sensor we measure the voltage output of the sensor and record the data. Fig.~\ref{Fig:sysResDifTrial} shows the system responses for 5 different trials. We wait between each trial until the initial voltage reading of the sensor drops to about~1 volts. Although it is extremely difficult to find the exact cause of deviations between trials, some likely causes are:  the spray, which is not precise enough to spray the same amount of alcohol for each trial; the flow, which can be turbulent; the sensor, which can be noisy; and other environmental factors such as random flows within the room. 

\begin{figure}[!t]
 \centerline{\resizebox{1.015\columnwidth}{!}{\includegraphics{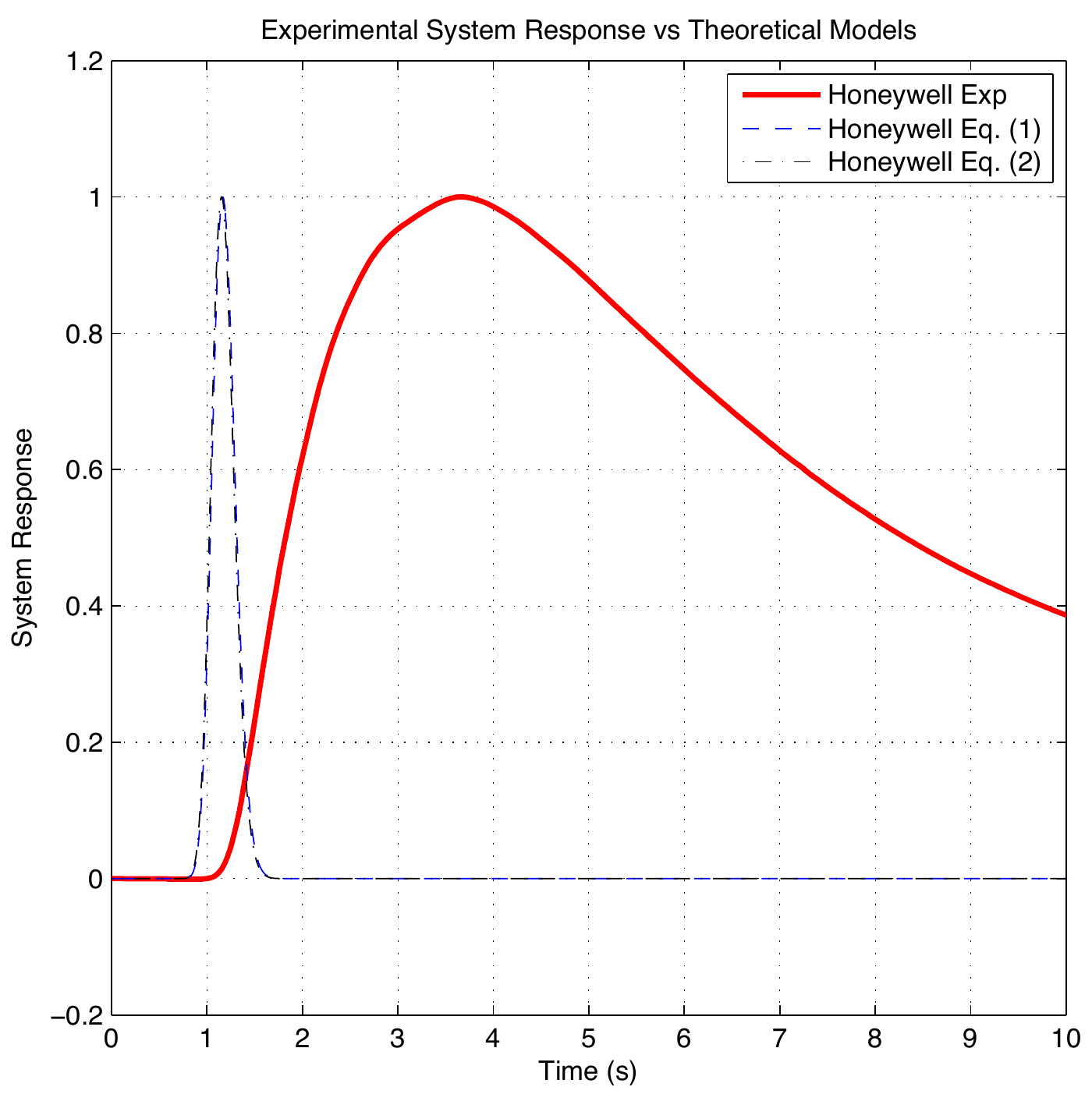}}}
  \caption{Comparison of the experimental data and theoretical models from previous works. The curves from the theoretical Equation (\ref{Eq:noAdsorb}) and (\ref{Eq:adsorb}) are very similar in this case and almost overlap. This follows because the speed of the flow is much greater than diffusion coefficient. Therefore, although the molecules are not absorbed by the sensor according to (\ref{Eq:noAdsorb}), they are moved away from the sensor by the flow.  }
  \label{Fig:oldTheorVsExp}
\end{figure}

\subsection{Previous Theoretical Models}
Based on theoretical models for concentration of particles in diffusion with drift, we assume that the spray and the sensor have the same height in the 3 dimensional space, and that the fan's flow is perfectly aligned with the line connecting the transmitter to the receiver\footnote{These assumptions can be easily satisfied through careful placement of the transmitter and the receiver}. Therefore, the problem reduces to a one dimensional Wiener process with drift. If we also assume that the sensor and the transmitter are perfect, and that the sensor does not absorb (or adsorb) the alcohol molecules (i.e. the alcohol molecules stay in the environment after detection), the impulse response at the receiver should be well approximated by~\cite{TIT_Andrew12}:
\begin{align}
\label{Eq:noAdsorb}
h_1(t) &= \frac{M}{\sqrt{4\pi D t}}\exp{\Big(-\frac{(d-vt)^2}{4Dt}\Big)}, 
\end{align}
where $M$ is the number of molecules released during the short burst, $D$ is the diffusion coefficient, $d$ is the separation distance between the transmitter and the receiver, $v$ is the average flow speed from the transmitter to the receiver, and $t$ is time. This is essentially the probability distribution of a Wiener process with drift, conditioned at a fixed distance $d$ and multiplied by the number of molecules $M$. If we assume that the alcohol molecules are absorbed (or adsorbed) by the sensor upon detection, then the problem would be equivalent to the first arrival time, and the impulse response would have a similar shape to the inverse Gaussian distribution~\cite{TIT_Andrew12} given by 
\begin{align}
\label{Eq:adsorb}
h_2 &= \frac{Md}{\sqrt{4\pi D t^3}}\exp{\Big(-\frac{(vt-d)^2}{4Dt}\Big)}. 
\end{align}

Although the number of molecules sprayed by the transmitter is not known (in fact it is random because each spray is not perfectly and precisely similar to previous sprays), based on theoretical results, it is expected that the sensor output should have a shape similar to the curves obtained from either (\ref{Eq:noAdsorb}) (in case the molecules are not absorbed by the sensor) or (\ref{Eq:adsorb}) (in case they are absorbed by the sensor).

\subsection{Models versus Experimental Results}
In this section we show that the previously published theoretical models described in the previous section do not match the experimental results obtained using the this tabletop platform. To demonstrate this, we separate the transmitter and the receiver by 225 cm. We use the Honeywell fan set on the high setting to generate flows. Table~\ref{tb:parameters} summarizes all the system parameters of this setup. The flow speed of the wind generated by the fan, which is tabulated in Table~\ref{tb:parameters} is measured using Pyle PMA82 digital anemometer.

\begin{table}[!t]
\caption{System Parameters.}
\begin{center}
\begin{tabular}{|c|c|}
\hline
Parameters & Values \\
\hline\hline
Spraying duration for each bit & 100 ms\\ \hline
Distance between a transmitter and a receiver & 225 cm \\ \hline
Approximated fan speed Honeywell & 190 cm/s \\ \hline
Diffusion coefficient of isopropyl alcohol & 0.0959 cm$^2$/s \\ \hline
Temperature (room temperature) & 25 $^{\circ}$C = 298 K \\ \hline
\end{tabular}
\end{center}
\label{tb:parameters}
\end{table}

If these parameters are used in the theoretical Equations (\ref{Eq:noAdsorb}) and (\ref{Eq:adsorb}), the system response can be calculated. Because the number of particles released by the transmitter is not known, we assume $M=1$ and then normalize the plots by dividing them by their respective maximums. Similarly the system responses obtained from experimental results is normalized with its maximum. By normalizing the plots, we compare only the shape of the theoretical results with the shape of the experimental results. For our experimental system response, we average the response of 12 different experimental trials to produce a single plot. Moreover, the initial voltage is subtracted from the system response to effectively zero the starting voltage. 

As shown in Fig.~\ref{Fig:oldTheorVsExp}, we can see that the experimentally obtained response has a much wider peak width, and longer tail compared to theoretical predictions. The difference between the theoretical results and the observed system response is because of many assumptions made in the derivation of the theoretical results. For example, the flow is assumed to be perfectly laminar and the sensor are assumed to be perfect at detection of concentration. These assumptions do not hold for our experimental platform. Therefore, in the next section we try to derive more realistic theoretical models based on the observed experimental data.

\begin{figure*}[!ht]
        \centering
        \begin{subfigure}{\textwidth}
                \centering
                \includegraphics[width=\textwidth]{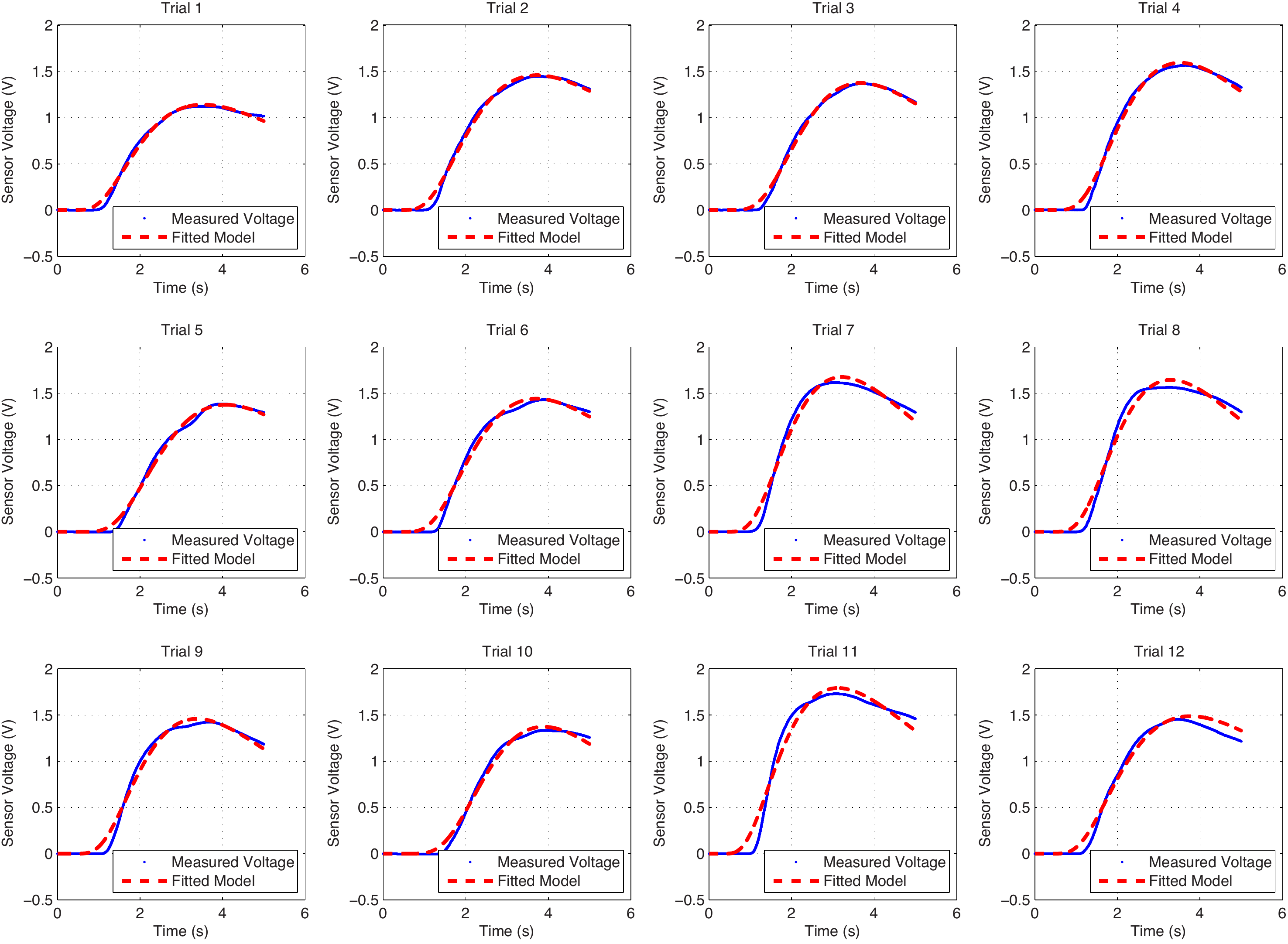}
                \caption{Sensor measurements and the fitted model for set of 12 trials. The measurements are fitted with model $M_1$.}
                \label{Fig:M1}
        \end{subfigure}%
        \\
        
        \begin{subfigure}{.7\textwidth}
                \centering
                \includegraphics[width=\textwidth]{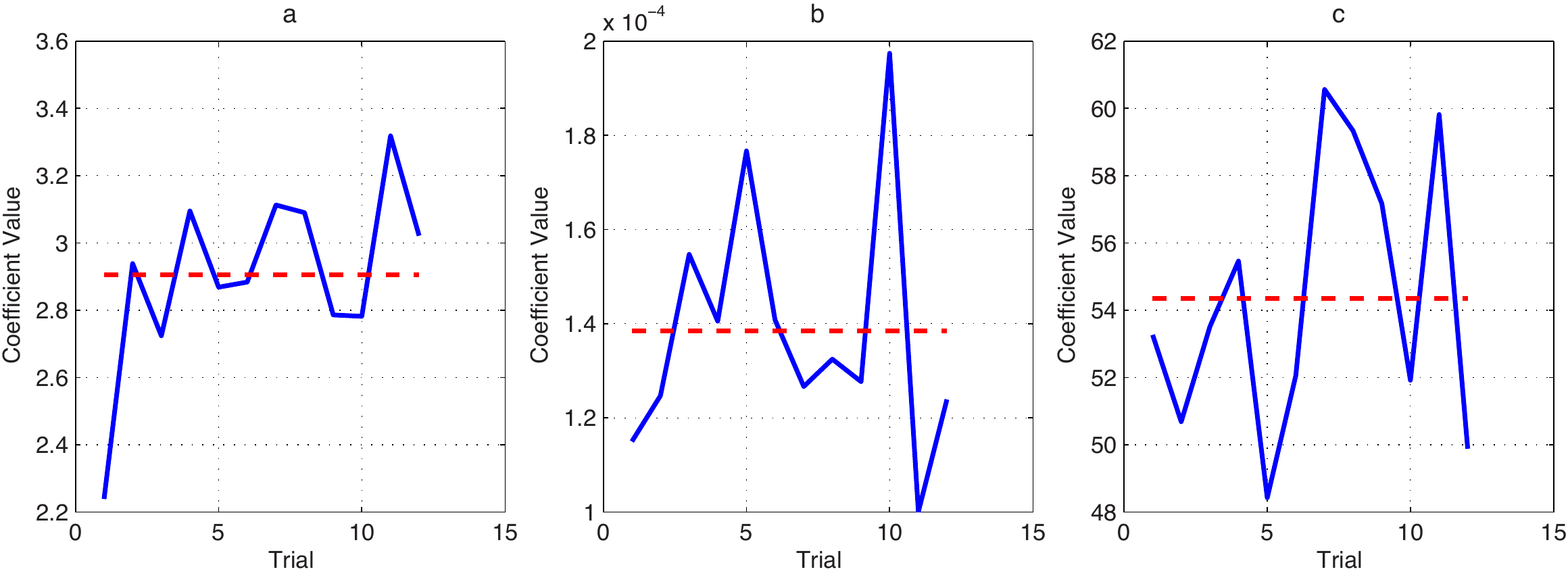}
                \caption{The coefficients' variation for 12 trials. The dashed red line is the mean value of each coefficient.}
                \label{Fig:M1coef}
        \end{subfigure}
        ~ 
        \caption{The curve fitting results.}\label{Fig:fitting}
\end{figure*}
\section{Realistic Models}
\label{Sec:Realistic Models} 
In this section we use our experimental data to derive a more realistic theoretical model for our platform. First, we find likely causes of the deviation from the theoretical results. In particular, two system components can have a huge effect on the system response: the sensor, and the flow. The previously published  channel models assume a perfectly laminar flow, as well as perfect and instantaneous detection at the sensor. These assumptions do not hold for this experimental platform. 

All metal-oxide sensors, have a response time and a recovery time~\cite{boc10}. The response time is the time it takes for the sensor to respond to a sudden change in concentration. The recovery time is the time it takes for the sensor to drop to its initial voltage after a sudden change in concentration. The concentration function with respect to time is therefore expanded because of the response and recovery times. To compensate for this effect, the system response function in~(\ref{Eq:noAdsorb}) and~(\ref{Eq:adsorb}) must be scaled in time by a factor of $\alpha$ as $h_1(\alpha t)$ and $h_2(\alpha t)$, where $0<\alpha<1$. 

Another factor that affects the system response is the flow. Previous channel models have assumed the flow to be perfectly laminar and uniform. This is, however, not the case for our platform. The high wind speeds generate turbulences within the flow. Moreover, the Honeywell fan's blades can create pockets of air pressure that can result in more turbulent flows. Fortunately, Fick's law of diffusion can still be applied to turbulent flows with a correction term added to the diffusion coefficient~\cite{guh08}. Therefore, a correction must be made to the diffusion coefficient~$D$ in~(\ref{Eq:noAdsorb}) and~(\ref{Eq:adsorb}).

The final factor we consider is the flow speed. Although we measure the wind speed generated by our fans, the alcohol droplets in the spray stream may be travelling at a slower average speed because of their weight and air friction. Therefore, a third correction is needed in~(\ref{Eq:noAdsorb}) and~(\ref{Eq:adsorb}) for the average flow velocity $v$. 

Considering these three effects, we propose two new models based on~(\ref{Eq:noAdsorb}) and~(\ref{Eq:adsorb}), 
\begin{align}
M_1(t) &= \frac{a}{\sqrt{t}}\exp{\Big(-b\frac{(d-ct)^2}{t}\Big)}, \label{Eq:M1} \\ 
M_2(t) &= \frac{a}{\sqrt{t^3}}\exp{\Big(-b\frac{(ct-d)^2}{t}\Big)}, \label{Eq:M2}
\end{align}
where $a$, $b$, and $c$ are corrected constants. The corrected constant $a$ contains the scaling factor $\alpha$ from the sensor respond and resume times, and the correction to the diffusion coefficient because of turbulent flow. The corrected constant $b$ contains the correction to diffusion coefficient because of turbulent flow and scaling factor $\alpha$. Finally, the corrected coefficient $c$ contains the correction to the average flow speed as well as the scaling factor $\alpha$.

\subsection{Estimating the Coefficients}

Finding the value of these proposed corrections can be very challenging. Therefore, we use the experimental data from our platform to estimate the value of these corrections. To do this we place the transmitter and the receiver 225 cm apart. We place the sensor, the spray and the fans at the same height, with the fan blowing in the direction of the line connecting the spray to the sensor.  We measure and record the end-to-end system impulse response to a very short spray burst of 100 ms during 12 different experimental trials. 

To estimate the coefficients of models $M_1$ and $M_2$, let $M_i(t_k,\mathbf{p})$ $i \in \{1,2\}$ be the corresponding model (i.e. model $M_1$ or $M_2$) at the sampled time instance $t_k$ with parameter vector $\mathbf{p}=[a,b,c]^T$, where $a$, $b$, and $c$ are the three coefficients for each model. We use nonlinear least square curve fitting for estimating the coefficients. Assuming that there are $N$ points in each sensor measurement and that each point in the measurement is represented by a function $m(t_k)$ $k \in \{1, 2,\cdots, N\}$, the coefficient estimation problem can be formulated as
\begin{align}
	 \min_{\mathbf{p}} \sum_{k=1}^N \big( m(t_k)-M_i(t_k,\mathbf{p}) \big)^2.
\end{align}
This problem can then be solved using iterative algorithms such as Levenberg-Marquardt algorithm~\cite{seb03}.

To perform the nonlinear least square estimation using Levenberg-Marquardt algorithm, MATLAB's curve fitting function \texttt{fit()} is used for coefficients estimation of each experimental trial. Because in~\cite{far13} it was demonstrated that only the first few seconds of the impulse response is typically used in practice for information transmission, we only use the first 5 seconds of sensor measurements for curve fitting.  Fig.~\ref{Fig:fitting} shows the results where the Honeywell fan on the high setting is used for flow generation, and model $M_1$ is used for curve fitting. In Fig.~\ref{Fig:M1}, we can see that the fitted model resembles the obtained results much more accurately compared to Fig.~\ref{Fig:oldTheorVsExp}. Fig.~\ref{Fig:M1coef} shows the plot of each coefficient value for each trial. The dashed red line indicates the mean value of each coefficient.

For the goodness of fit measure, we use the root mean square error (RMSE) between the fitted model and the experimentally observed system responses. We also use the variance-to-mean ratio (VMR) as a goodness of fit measure. If this ratio is greater than one, then the resulting coefficient is not a good fit. If this ratio is less than 1, then the coefficient is a good fit. Table~\ref{Table:SISOcoefs} summarize the result for both model $M_1$ and model $M_2$ given in (\ref{Eq:M1}) and (\ref{Eq:M2}), respectively. In the table the mean RMSE is the average RMSE across all 12 experimental trials.

From the results we can see that model $M_1$ has a better VMR, while model $M_2$ has a slightly better RMSE. Generally, because model $M_1$ has a lower VMR, the coefficients are more consistent between different experimental trials. Therefore, model $M_1$ may be more effective at consistently modelling the end-to-end system impulse response.

\begin{figure}[!t]
 \centerline{\resizebox{\columnwidth}{!}{\includegraphics{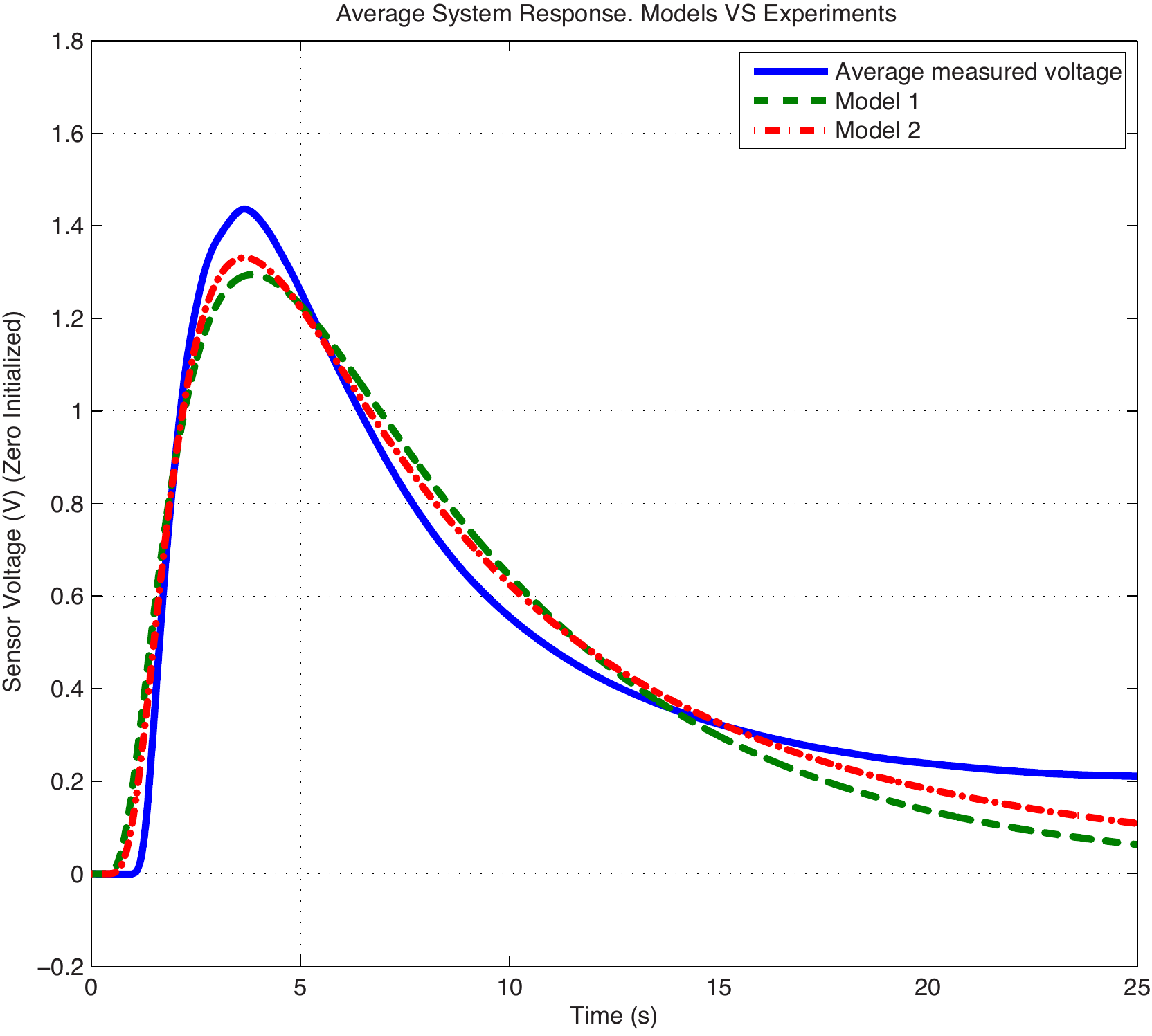}}}
  \caption{Average system response of experimental observations and fitted models.}
  \label{Fig:avgTheorVsExp}
\end{figure}

To further compare the proposed models to the experimental results, we average the system response from all 12 trials to generate the averaged experimental system response. We also use the mean of the coefficients across all 12 trials in each model to generate the corresponding system response. The results are shown in Fig.~\ref{Fig:avgTheorVsExp}. Based on the results we can see that both new models capture the average system response of the test bed platform much more accurately compared to old models.

\begin{table}[!h]
\caption{The obtained coefficients for each model.}
\begin{center}
\begin{tabular}{ |l|l|l|l|l|l|l|l|l|l|}
  \hline
    \multicolumn{5}{|c|}{Model $M_1$} \\

  \hhline{=====}
     & Mean & Variance& Variance/Mean & Mean RMSE \\ \hline 

  a & 2.9050 & 0.0672 & 0.0231 &\\
  b & 1.3839$\times10^{-4}$ & 6.6117$\times10^{-10}$ & 4.7775$\times10^{-6}$ & 0.0539\\
  c & 54.3405 & 15.3455 & 0.2824 &\\
  \hhline{=====}
    \multicolumn{5}{|c|}{Model $M_2$} \\
  \hhline{=====}
     & Mean & Variance& Variance/Mean & Mean RMSE \\ \hline

  a & 15.3909 & 4.2150 & 0.2739 &\\
  b & 1.6$\times10^{-4}$ & 6.9109$\times10^{-10}$ & 4.31$\times10^{-6}$ & 0.0501\\
  c & 35.3136 & 29.5543 & 0.8369 &\\
  \hline
\end{tabular}
\end{center}
\label{Table:SISOcoefs}
\end{table}

\section{System's Nonlinearity}
\label{nonliearity}
In~\cite{far13}, it was demonstrated that the platform has a nonlinear system impulse response. This nonlinearity property, however, was not investigated in great detail. Although it is extremely difficult to find the exact cause of nonlinearity (some likely causes are imperfect receiver, and transmitter and turbulent flows), in~\cite{far13} the nonlinearity property was demonstrated through measurements. In this section, we first verify the nonlinearity property through systematic experimentation using two transmitters and a single receiver. We then represent the nonlinearity as noise and find the underlying distribution for this noise by employing the developed models in the previous section.
\begin{figure}[!t]
	\begin{center}
       \includegraphics[width=\columnwidth]{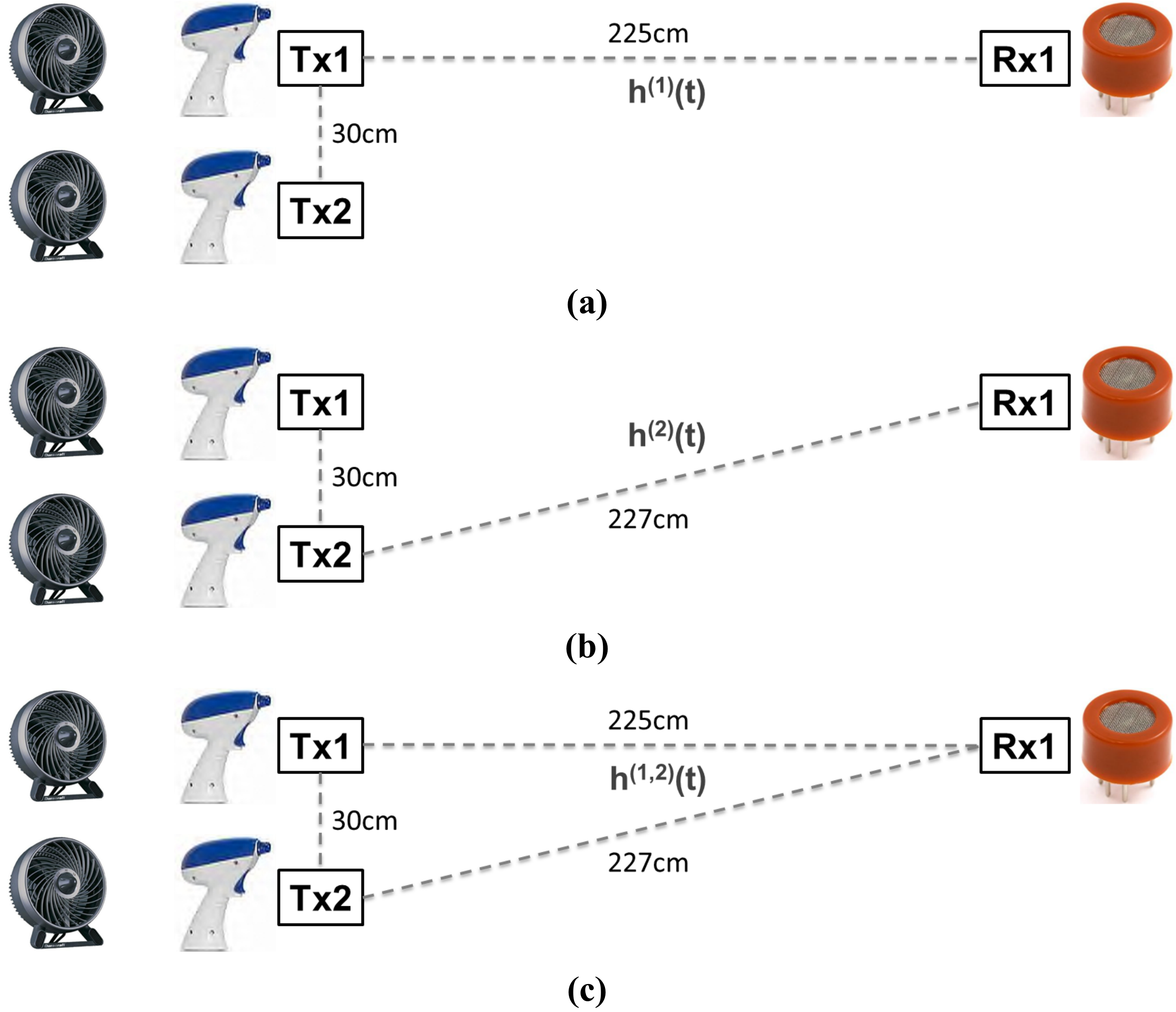}
	\end{center}
    \caption{The setup used to demonstrate the nonlinearity.}\label{Fig:MISO}
\end{figure}
\begin{figure}[!t]
 \centerline{\resizebox{\columnwidth}{!}{\includegraphics{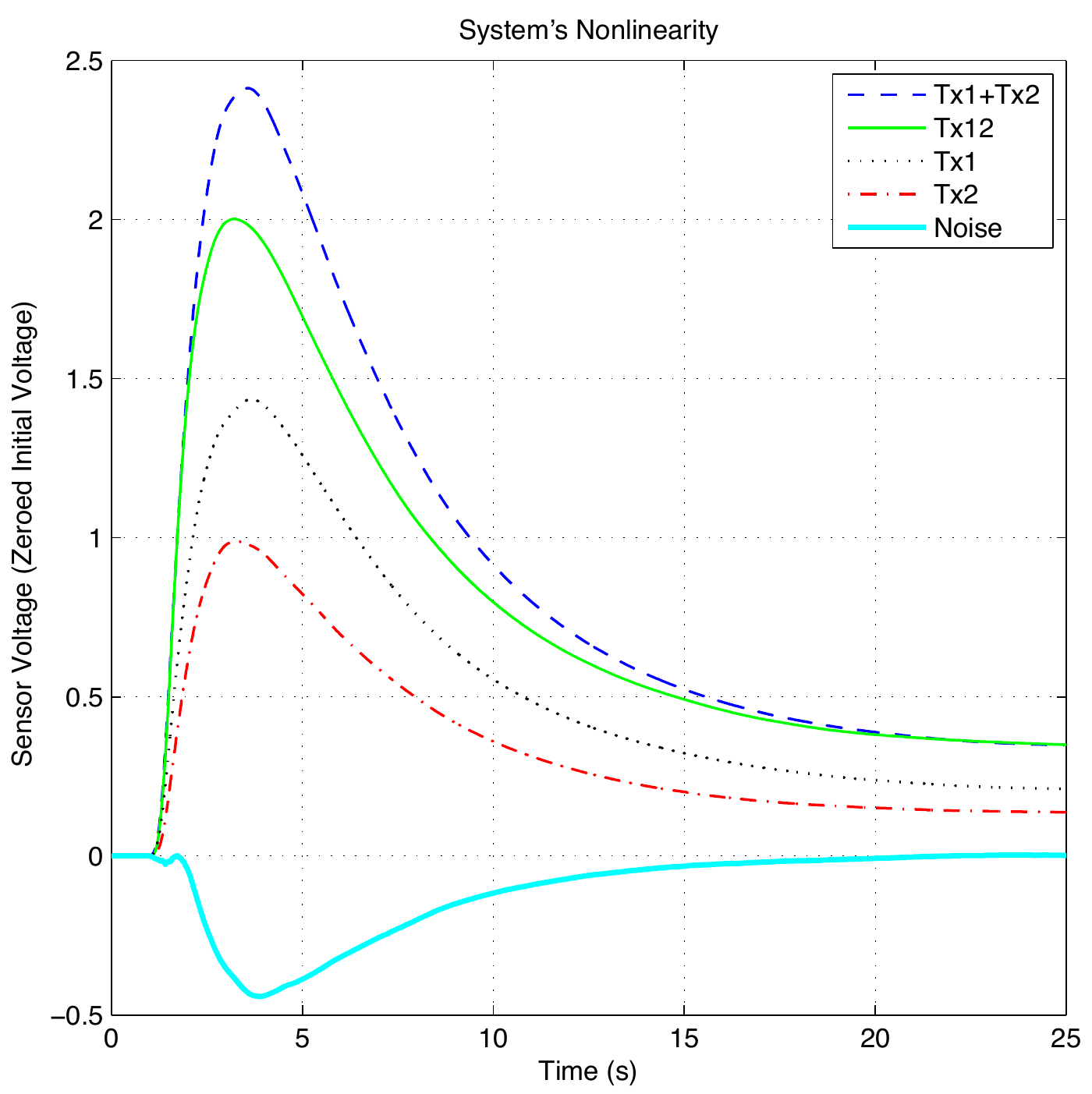}}}
  \caption{Representation of system's nonlinearity.}
  \label{Fig:sysNonLinear}
\end{figure}
To study the nonlinearity, we use two transmitters and one receiver. Each transmitter has its own Honeywell fan. The transmitters are 30 cm apart, and the receiver is directly in front of the first transmitter separated by 225 cm. Fig.~\ref{Fig:MISO} summarizes this setup. Let $h^{(1)}(t)$ and $h^{(2)}(t)$ be the end-to-end system impulse response, when transmitter 1 and 2 spray a short burst of 100 ms in duration, respectively. Moreover, let $h^{(1,2)}(t)$ be the end-to-end system impulse response when both transmitters spray a short burst of 100 ms in duration simultaneously. If the system is linear, then we have
\begin{align}
	h^{(1,2)}(t) = h^{(1)}(t) +  h^{(2)}(t).
\end{align}  
As shown in Fig.~\ref{Fig:sysNonLinear} this property does not hold. The system responses for $h^{(1)}(t)$ (Tx1), $h^{(2)}(t)$ (Tx2), and $h^{(1,2)}(t)$ (Tx12) are generated by averaging 12 different trials to reduce the noise that may be introduced by other processes such as random flow patterns in the room. As can be seen the $h^{(1,2)}(t)$ (Tx12) plot and the $h^{(1)}(t) +  h^{(2)}(t)$ (Tx1 + Tx2) plot are not equal. This verifies the claim by~\cite{far13} that the system tends to be nonlinear.

Although there may be many contributors to the end-to-end system nonlinearity, based on these results, we believe one of the most important contributors is the sensor. In essence the sensor may act as a nonlinear filter. In the next section we try to model this nonlinearity as a noise process. Through this formulation we can counter the effects of this nonlinear filter.   

\subsection{Modelling the Nonlinearity as Noise}

The nonlinearity in the system can be modelled as noise. Because the functions $h^{(1)}(t)$, $h^{(2)}(t)$, and $h^{(1,2)}(t)$ are random processes, the nonlinearity noise can be represented as a random process given by
\begin{align}
     \label{eq:additiveNoise}
	 h^{(1,2)}(t) =  h^{(1)}(t) +  h^{(2)}(t) + n(t),
\end{align}
where $n(t)$ is the noise process. To find the underlying model for the noise process using experimental measurements, we rewrite (\ref{eq:additiveNoise}) as  
\begin{align}
     \label{eq:noise}
	 n(t) = h^{(1,2)}(t) - h^{(1)}(t) -  h^{(2)}(t).
\end{align}
The expected value of the noise process is then given by:
\begin{align}
	 E[n(t)] = E[h^{(1,2)}(t)] - E [h^{(1)}(t)] - E[h^{(2)}(t)].
\end{align}
In Fig.~\ref{Fig:sysNonLinear} the expected value of the noise process is represented by the black dashed plot. The noise process in this case is therefore nonstationary, because its expected value in not a constant in time. 

To simplify the noise model, we use the mathematical models we derived in the previous sections. 
To find the coefficients of this model we measure $h^{(1)}(t)$, $h^{(2)}(t)$, and $h^{(1,2)}(t)$ using 36 different trials (12 distinct trials for each case). Let  $h^{(1)}_i(t)$, $h^{(2)}_j(t)$, and $h^{(1,2)}_k(t)$ ($i,j,k=1,2,\ldots,12$) be the results of each trial. Using the MATLAB \texttt{fit()} function, we estimate the coefficients of model $M_1$ and $M_2$ based on the observed data from each trial.  Again we use the first 5 seconds of sensor measurements for curve fitting, since in practice this would be the information carrying interval. Table~\ref{Table:MISOcoef} shows the average value of each coefficient across different trials.   
\begin{table}[!t]
\caption{The obtained mean coefficients of model function.}
\begin{center}
\begin{tabular}{ |l|l|l|l|l|l|l|l|l|}
  \hline
  \multicolumn{4}{|c|}{Model $M_1$} \\
  \hhline{====}
Coefficient & $h^{(1)}$ 			&  $h^{(2)}$   			& $h^{(1,2)}$ \\ \hline
$a$			&2.905					& 1.9815				&3.9737    \\
$b$			&1.3839$\times10^{-4}$	&1.5605$\times10^{-4}$	&1.3474$\times10^{-4}$ \\
$c$			&54.3405				&59.4961				&58.669 \\
\hhline{====}
  \multicolumn{4}{|c|}{Model $M_2$} \\
\hhline{====}
Coefficient & $h^{(1)}$ 			&  $h^{(2)}$   			& $h^{(1,2)}$ \\ \hline
$a$			& 15.3909				& 12.6246				& 18.7617   \\
$b$			& 1.6035$\times10^{-4}$	& 1.5119$\times10^{-4}$	& 1.5885$\times10^{-4}$ \\
$c$			& 35.3137				& 28.9746				& 40.7749 \\
\hline
\end{tabular}
\end{center}
\label{Table:MISOcoef}
\end{table} 

From this table it can be seen that coefficients $b$ and $c$ do not differ by more than about 10 to 15 percent. Moreover, the separation distance between the transmitters 1 and 2, and the receiver are almost similar. However, coefficient $a$ changes significantly, depending on which transmitters are spraying. This is consistent with precious theoretical works, where system linearity is assumed. Assuming that the coefficients $b$ and $c$ are similar across different trials and different transmitter sprays, (\ref{eq:noise}) becomes
\begin{align}
     n(t) &= h^{(1,2)}(t) - h^{(1)}(t) -  h^{(2)}(t), \nonumber  \\
     n_{M_1}(t) &\approx \frac{a_{M_1}^{(1,2)}}{\sqrt{t}}\exp{\Big(\frac{-b(d-ct)^2}{t}\Big)} \\
     &- \frac{a_{M_1}^{(1)}}{\sqrt{t}}\exp{\Big(\frac{-b(d-ct)^2}{t}\Big)}  \nonumber\\
     &-\frac{a_{M_1}^{(2)}}{\sqrt{t}}\exp{\Big(\frac{-b(d-ct)^2}{t}\Big)},  \nonumber \\
     n_{M_1}(t) &\approx \frac{N_{M_1}}{\sqrt{t}}\exp{\Big(\frac{-b(d-ct)^2}{t}\Big)}, \label{eq:noiseSimpM1} \text{~and}    \\
	 N_{M_1} &= a_{M_1}^{(1,2)} - a_{M_1}^{(1)} -  a_{M_1}^{(2)},
\end{align}
where $a_{M_1}^{(1)}$, $a_{M_1}^{(2)}$, and $a_{M_1}^{(1,2)}$ are the first coefficients of model $M_1$ fitted to $h^{(1)}(t)$, $h^{(2)}(t)$, and $h^{(1,2)}(t)$, respectively, and $N_{M_1}$ is the simplified noise model. Using the same procedure a simplified noise model can be generated based on model $M_2$ as 
\begin{align}     
     n_{M_2}(t) &\approx \frac{N_{M_2}}{\sqrt{t^3}}\exp{\Big(\frac{-b(ct-d)^2}{t}\Big)}, \label{eq:noiseSimpM2} \text{~and}  \\
	 N_{M_2} &= a_{M_2}^{(1,2)} - a_{M_2}^{(1)} -  a_{M_2}^{(2)},
\end{align}
where $a_{M_2}^{(1)}$, $a_{M_2}^{(2)}$, and $a_{M_2}^{(1,2)}$ are the first coefficients of model $M_2$ fitted to $h^{(1)}(t)$, $h^{(2)}(t)$, and $h^{(1,2)}(t)$, respectively, and $N_{M_2}$ is the simplified noise. Using this method the noises becomes $N_{M_1}$ and $N_{M_2}$ become random variables.

To find underlying probability distribution of $N_{M_1}$ and $N_{M_2}$, we estimate the first coefficient ($a$ coefficients) of each model for each $h^{(1)}_i(t)$, $h^{(2)}_j(t)$, and $h^{(1,2)}_k(t)$ ($i,j,k=1,2,\ldots,12$) from our experimental trials. In these estimations we assume the value of coefficients $b$, $c$ and $d$ are constant, and use the average value of $b_{M_1} = 1.4306\times 10^{-4}$ and $c_{M_1} = 57.5018$ when model $M_1$ is used, and values of $b_{M_2} = 1.57\times 10^-4$ and $c_{M_2} = 35.021$ when model $M_2$ is used. These values are obtained by averaging the corresponding row in Table~\ref{Table:MISOcoef}. We also assume the distance to the receiver for both transmitters is $d=225$. From the obtained coefficients we generate 1728 ($12^3$) different noise samples using
\begin{align}
	 N_{M_1}[i,j,k] = a_{M_1}^{(1,2)}[k] - a_{M_1}^{(1)}[i] -  a_{M_1}^{(2)}[j], \label{eq:NoiseSampGenM1} \\
	 N_{M_2}[i,j,k] = a_{M_2}^{(1,2)}[k] - a_{M_2}^{(1)}[i] -  a_{M_2}^{(2)}[j], \label{eq:NoiseSampGenM2} 
\end{align}
\begin{figure}[t]
 \centerline{\resizebox{\columnwidth}{!}{\includegraphics{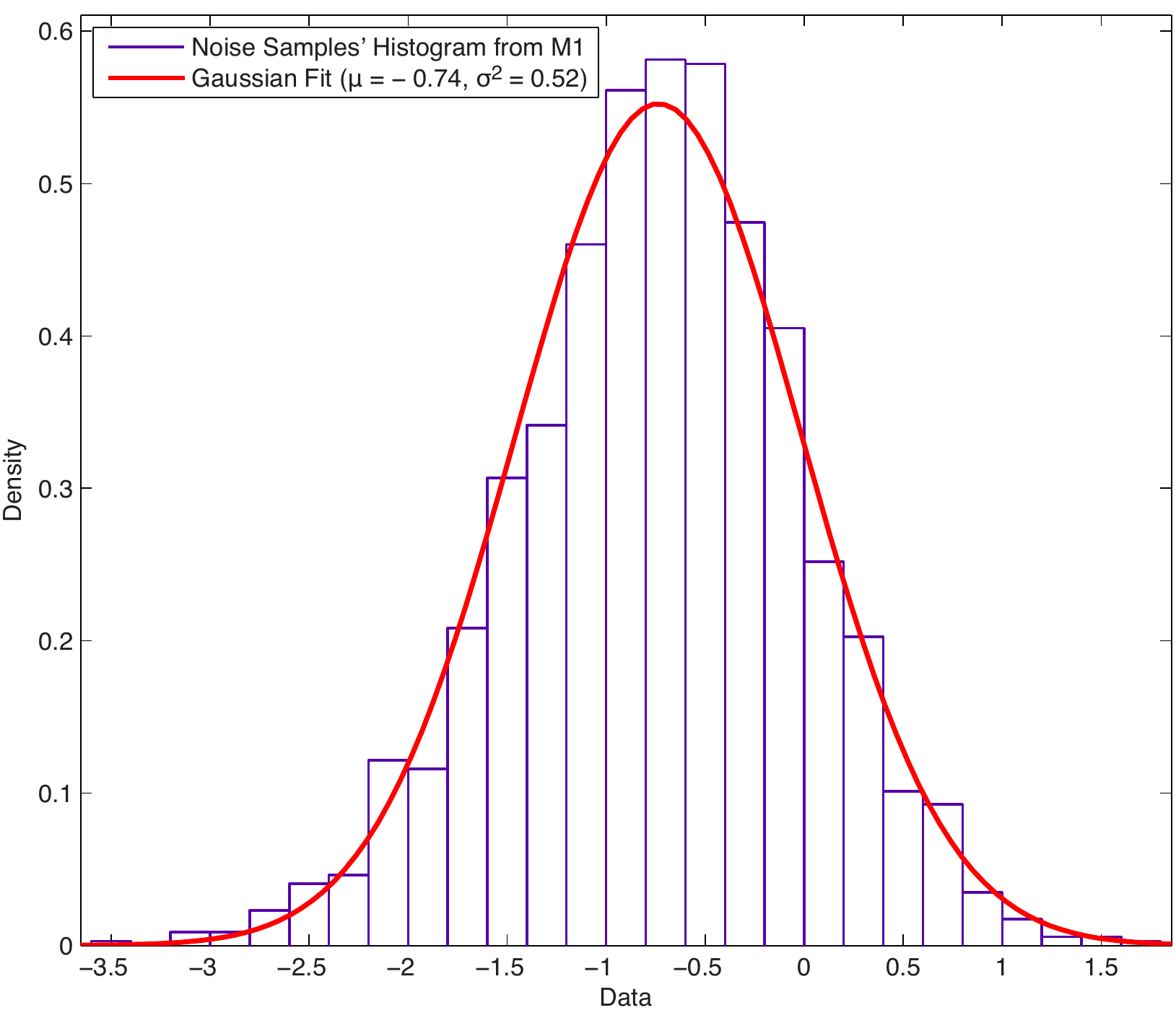}}}
  \caption{Histogram of the noise samples based on model $M_1$ and fitted Guassian probability density.}
  \label{Fig:noiseHistM1}
\end{figure}

\begin{figure}[t]
 \centerline{\resizebox{\columnwidth}{!}{\includegraphics{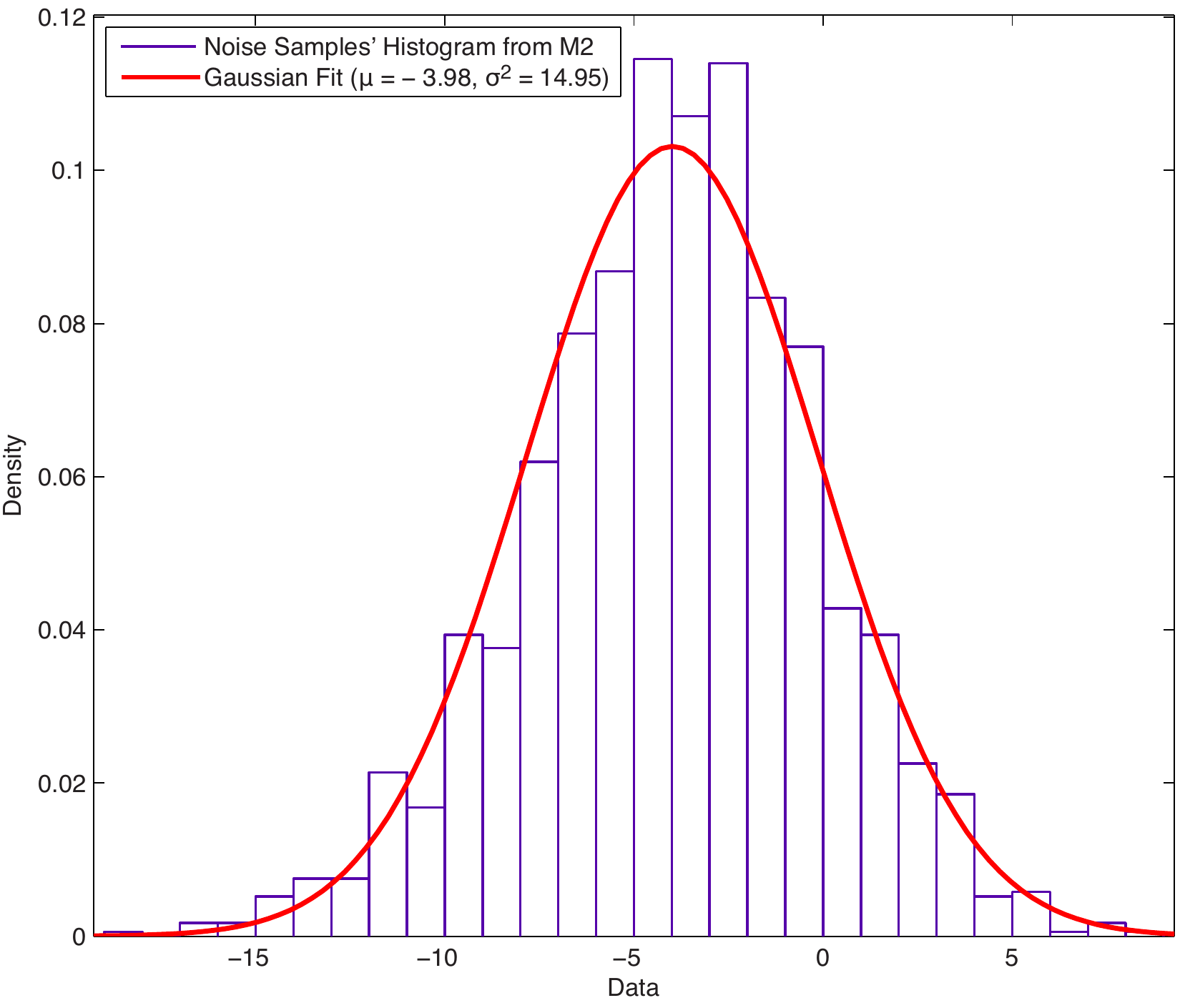}}}
  \caption{Histogram of the noise samples based on model $M_2$ and fitted Guassian probability density.}
  \label{Fig:noiseHistM2}
\end{figure}

Figs.~\ref{Fig:noiseHistM1} and~\ref{Fig:noiseHistM2} show the histogram of the of the sample noises generated using Equations~\ref{eq:NoiseSampGenM1} and~\ref{eq:NoiseSampGenM2}, respectively. As can be seem the results are close to Gaussian. Therefore, although the exact distribution is not known Gaussian assumption is favourable~\cite{sto11, par12}. The Gaussian fit plot is generated using the mean and the variance of the sample. The mean of the sample for samples generated based on model $M_1$ is $\mu=-0.7356$ and the variance of the sample is $\sigma^2=0.5214$. The mean and variance for the samples generated using model $M_2$ are $\mu=-3.9811$ and $\sigma^2=14.9589$. The samples generated based on model $M_2$ have a much larger variance because of the lager VMR of this model compared with model $M_1$.

\subsection{Results}
\label{Sec:results}
To validate our noise estimation model of nonlinearity, we generate two random noise processes using
\begin{align}
	 n_{M_1}(t) &\approx \frac{N_{M_1}}{\sqrt{t}}\exp{\Big(\frac{-b_{M_1}(d-c_{M_1}t)^2}{t}\Big)}, \label{eq:noiseProcM1} \\
	 n_{M_2}(t) &\approx \frac{N_{M_2}}{\sqrt{t^3}}\exp{\Big(\frac{-b_{M_2}(c_{M_2}t-d)^2}{t}\Big)} \label{eq:noiseProcM2}
\end{align}
where $N_{M_1}$ is the Gaussian random variable with mean $\mu=-0.7356$ and variance $\sigma^2=0.5214$, $b_{M_1} = 1.4306\times 10^{-4}$ (average value of the corresponding row in Table~\ref{Table:MISOcoef}), $c_{M_1} = 57.5018$ (average value of the corresponding row in Table~\ref{Table:MISOcoef}), and $d=225$ is the separation distance between the transmitter and the receiver. Similarly $N_{M_2}$ is the Gaussian random variable with mean $\mu=-3.9811$ and variance $\sigma^2=14.9589$, $b_{M_2} = 1.57\times 10^-4$ and $c_{M_2} = 35.021$. Using this noise process, 144 different samples for when both transmitters spray is generated using the sensor measurements from a single transmitter spray data using
\begin{align}
	 \hat{h}_{i,j}^{(1,2)} = h^{(1)}_i(t)+h^{(2)}_j(t)+n(t),
\end{align}
where $\hat{h}_{i,j}^{(1,2)}$ is the estimated sample, $h^{(1)}_i(t)$, and $h^{(2)}_j(t)$, are sensor measurement from 12 different trials ($i,j=1,2,\ldots,12$), and $n(t)$ is the noise process generated either using model $M_1$ (i.e. Equation~\ref{eq:noiseProcM1}) or model $M_2$ (i.e. Equation~\ref{eq:noiseProcM2}).

\begin{figure}[!t]
 \centerline{\resizebox{\columnwidth}{!}{\includegraphics{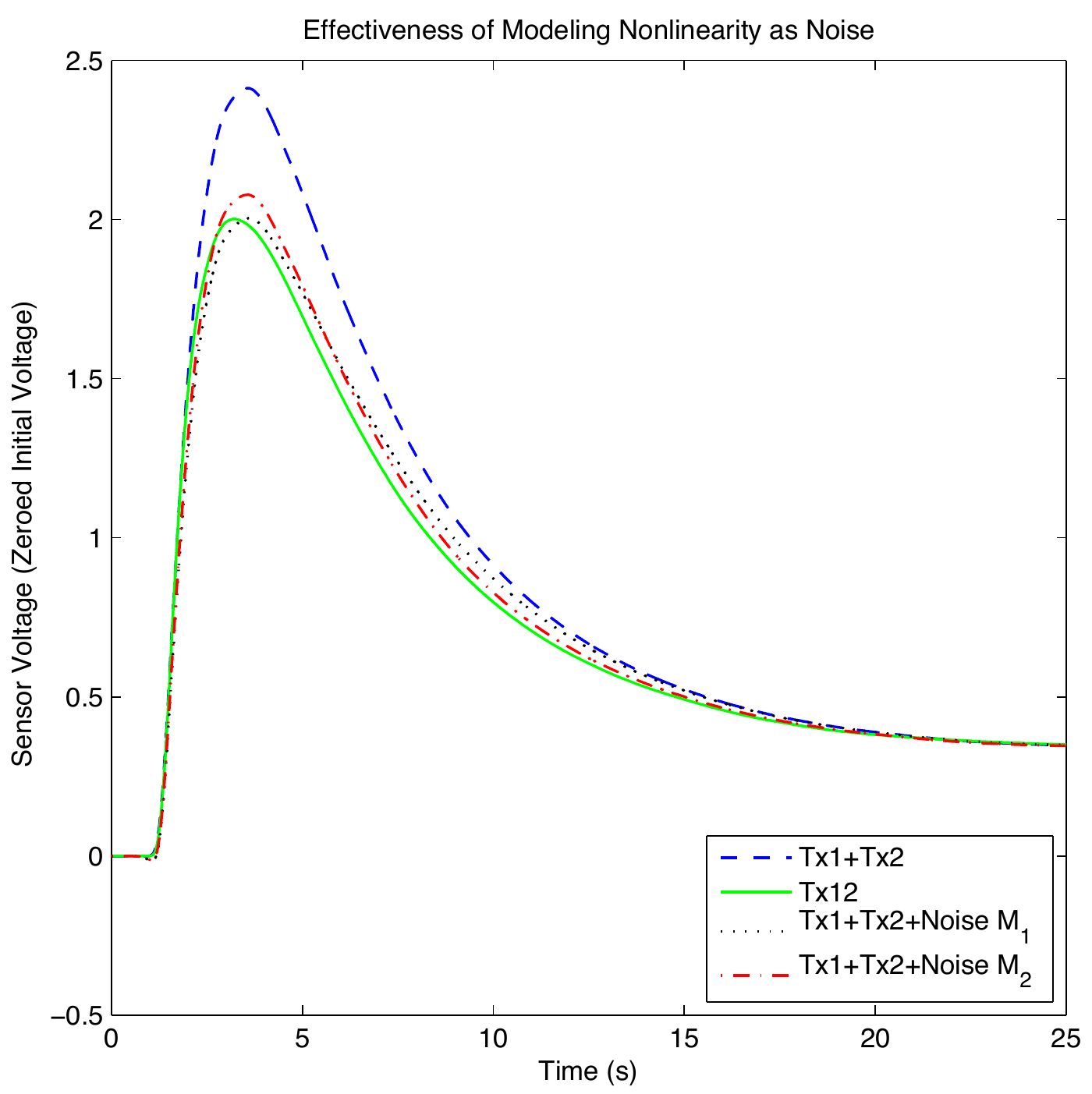}}}
  \caption{Effectiveness of modeling the nonlinearity as noise.}
  \label{Fig:results}
\end{figure}

Fig.~\ref{Fig:results} shows the results. The Tx12 plot shows the the average system response $h^{(1,2)}$ when both transmitters spray (averaged across 12 different trials). The Tx1+Tx2 plot shows the average system response for $h^{(1)}(t) + h^{(2)}(t)$, the Tx1+Tx2+Noise M$_1$ plot shows the average $\hat{h}^{(1,2)}$ across the 144 samples when the noise model is based on $M_1$, and the Tx1+Tx2+Noise M$_2$ plot shows the red plot shows the average $\hat{h}^{(1,2)}$ across the 144 samples when the noise model is based on $M_2$. As can be seen from the plot, both noise models that are presented can effectively represent the nonlinearity which is present in the system. This is a significant result since the system can now be represented as a linear model with noise. The noise estimation is fairly accurate. 

\section{Conclusions}
\label{Sec:Conc}
In this work we considered the tabletop molecular communication platform that was recently presented in~\cite{far13}, and is capable of transmitting short text messages across a room. Since well-known theoretical models for impulse response of  molecular communication with drift do not match with experimental data obtained using the platform, we suggested a new realistic channel model from experiments. We achieved this by systematically introducing correction factors the previously published models for impulse response and estimating the value of the correction factors based on experimental observations. We then used the derived models to study the nonlinearity of this tabletop platform. 

First, we systematically demonstrated that the platform is indeed nonlinear. We then modeled the nonlineariry as noise, and used the derived models with correction factors to show that with some simplifying assumptions this noise can be represented as Gaussian noise, which is known to be a good approximation for cases when there are no known models for the noise~\cite{sto11, par12}. We then evaluated the effectiveness of using this noise model for representing the nonlinearity and demonstrated that it can be a good model for this platform. By representing the
nonlinearity as a linear system with with Gaussian noise, a large body of theoretical work can now be applied to this platform.

For the future work, we will consider applying multiple-input multiple-output techniques to this platform to increase the achievable data rates. Eventually, previous work based on the theoretical model will be revisited with our new realistic model. For instance, in~\cite{TNBS_Kim13}, we can further optimize the symbol intervals by applying a new channel model.

\bibliographystyle{IEEEbib}

\bibliography{references_JSAC2013}


\begin{biography}{Nariman Farsad}
received his B.Sc. and M.Sc. degrees in computer science and engineering from York University, Toronto, ON, Canada in 2007 and 2009, respectively.  He is currently a Ph.D. student with the Department of Electrical Engineering and Computer Science at York University. He is interested in studying molecular and biological communication systems through an information-theoretic lense.
\end{biography}

\begin{biography}{Na-Rae Kim}
(S'12) received her B.S. degree in Chemical Engineering from Yonsei University, Korea in 2011. She is now with the School of Integrated Technology at the same university and is working toward the Ph.D. degree. She was an exchange student at University of California, Irvine in USA in 2009. 

Ms. Kim is the recipient of the travel grant from the IEEE International Conference on Communications in 2012 and the Gold Prize in the 19th Humantech Paper Award. 
\end{biography}

\begin{biography}{Andrew Eckford}
(M'96 - S'97 - M'04) is originally from Edmonton,
AB, Canada. He received the B.Eng. degree in
electrical engineering from the Royal Military
College of Canada, Kingston, ON, Canada, in 1996,
and the M.A.Sc. and Ph.D. degrees in electrical
engineering from the University of Toronto, Toronto,
ON, Canada, in 1999 and 2004, respectively.
He is currently an Associate Professor of Computer Science and Engineering at York University,
Toronto.
Dr. Eckford is also the Chair of the IEEE ComSoc
Emerging Technical Subcommittee on Nanoscale, Molecular, and Quantum
Networking.
\end{biography}

\begin{biography}{Chan-Byoung Chae}
(S'06 - M'09 - SM'12) is an Assistant Professor in the School of Integrated Technology, College of Engineering, Yonsei University, Korea. He was a Member of Technical Staff (Research Scientist) at Bell Laboratories, Alcatel-Lucent, Murray Hill, NJ, USA from 2009 to 2011. Before joining Bell Laboratories, he was with the School of Engineering and Applied Sciences at Harvard University, Cambridge, MA, USA as a Post-Doctoral Research Fellow. He received the Ph. D. degree in Electrical and Computer Engineering from The University of Texas (UT), Austin, TX, USA in 2008, where he was a member of the Wireless Networking and Communications Group (WNCG).

Prior to joining UT, he was a Research Engineer at the Telecommunications R\&D Center, Samsung Electronics, Suwon, Korea, from 2001 to 2005. He was a Visiting Scholar at the WING Lab, Aalborg University, Denmark in 2004 and at University of Minnesota, MN, USA in August 2007. While having worked at Samsung, he participated in the IEEE 802.16e standardization, where he made several contributions and filed a number of related patents from 2004 to 2005. His current research interests include capacity analysis and interference management in energy-efficient wireless mobile networks and nano (molecular) communications. He serves as an Editor for the \textsc{IEEE Trans. on Wireless Communications}, \textsc{IEEE Trans. on Smart Grid}, and \textsc{Jour. of Comm. Networks}. He is also an Area Editor for the \textsc{IEEE Jour. Selected Areas in Communications} (nano scale and molecular networking). He is an IEEE Senior Member.

Dr. Chae was the recipient/co-recipient of the IEEE Signal Proc. Mag. Best Paper Award in 2013, the IEEE ComSoc AP Outstanding Young Researcher Award in 2012, the IEEE Dan. E. Noble Fellowship in 2008, the Gold Prize (1st) in the 14th/19th Humantech Paper Award, and the KSEA-KUSCO scholarship in 2007. He also received the Korea Government Fellowship (KOSEF) during his Ph. D. studies.
\end{biography}

\end{document}